# Controlling Thermal Emission by Parity-symmetric Fano Resonance of Optical Absorbers in Metasurfaces

*Xia. Zhang[1,2*], Zhen-guo. Zhang[1], Qiang. Wang[1], Shi-ning. Zhu[1] and Hui. Liu[1*]*

[1]National laboratory of solid state microstructures and school of physics, Collaborative Innovation Center of Advanced Microstructures, Nanjing University, Nanjing 210093, China

[2]School of Physics and Physical Engineering, Shandong Provincial Key Laboratory of Laser Polarization and Information Technology, Qufu Normal University, Qufu 273165, China

ABSTRACT In this study, we designed a metasurface based on Al/SiN/Al nano-sandwich optical absorbers. Parity-symmetric dark magnetic resonance modes, anti-symmetric bright magnetic resonance modes, and surface lattice modes are established simultaneously in such a meta-surface. By changing the structural parameters, the wavelength of the dark mode can be tuned to overlap with the lattice mode. Fano resonance is introduced by coupling between the dark and lattice modes. Angular-resolved thermal emission experiments showed that the thermal emission induced by Fano resonance exhibited a narrow wavelength spectrum and directional radiation angle. Fano-type thermal emitters proposed in this work can possibly be used as coherent infrared light sources in the future.

In recent years, research interest in thermal emissions in the infrared wavelength range has rapidly grown due to its important applications in thermophotovoltaic (TPV) devices[1], radiative cooling[2,3], incandescent sources[4], near-field heat transfer[5], and infrared spectroscopy[6]. Various structures and systems have been used to control thermal emissions, such as gratings[7], nanoantennas[8], photonic crystals[9], surface plasmons[10,11], metamaterials[12–15], and metasurfaces (MTSs)[16]. Control of the properties of thermal emission, such as the emission bandwidth[14,17], coherent properties[7] and dynamics switching[18–20], has been reported.

In plasmonics, researchers can design and fabricate various metallic nanostructures with different shapes and sizes to produce different plasmonic resonance. In complex systems, the tuning of the interactions between different plasmonic resonators can provide a flexible method for producing many interesting novel resonance modes[21]. These modes have significantly different far-field radiation properties[22]. Some modes have strong far-field radiation, which are defined as bright modes. However, some modes have weak far-field radiation, which are defined as dark modes. When bright and dark modes are coupled with each other, the well-known Fano resonance mode can be produced[23–25]. Compared with uncoupled single resonance modes, Fano resonance exhibits a higher Q factor and narrower bandwidth, which has applications in optical sensing and light-matter interactions.

Most types of plasmonic resonance are either electric resonance (ER) or magnetic resonance (MR). Typically, MR exhibits better optical absorption and local field enhancement than ER. Perfect absorber based on MR mode in metamaterials had been studied [26,27]. Compared with bulk metamaterials, 2D MTSs are more easily fabricated, and the optical losses can be suppressed significantly[28,29], and MTSs have been widely used in both basic and application researches[30-35]. Among different MTS resonance unit designs, metal/insulation/metal (MIM) sandwich resonators have very strong local MR modes and are easily fabricated. MIM resonators have been used in magnetic field enhancement[15,36], nonlinear generation[37], negative optical pressure[38] and perfect absorbers[26,27]. According to Kirchhoff's law, thermal radiation of a medium is proportional to its optical absorption. Thus, MIM can act as a good thermal emitter. Using the thermal radiation scanning tunneling microscopy technique, the electromagnetic field spatial distribution of individual MIM patch nanoantenna was revealed[39].

In recent years, symmetry analysis has played an increasingly important role in optical systems. Through designing optical modes with different symmetry, modes coupling can be manipulated[40], and many novel interesting optical properties, such as symmetry-protected topological modes[41–43] and PT-symmetric modes[44–46], can be introduced in optics. Usually, the far-field radiation properties of plasmonic modes are related to their symmetry. Therefore, through designing plasmonic resonators with different symmetries, we can obtain plasmonic resonators with different radiation properties. In this work, we designed and fabricated a magnetic MTS composed of a periodic MIM array. The MIM can have multiple MR modes at different wavelengths. The spatial symmetries of these modes were different. Some MRs were parity-symmetric dark modes and others were parity-antisymmetric bright modes. Simultaneously, the interference between the MIMs in the periodic array could produce the surface lattice resonance (SLR) mode, which exhibited very strong far-field radiation and was regarded as a bright mode. By changing the width of the MIM, we tuned the resonance frequency of the MR. In this way, the MR frequencies can overlap with the SLR frequencies, which introduces strong coupling between the MR and SLR modes. When a bright MR mode was coupled with an SLR mode, we obtained an ordinary bright mode. When a dark MR mode was coupled with an SLR mode, we produced Fano resonance. Compared with uncoupled MR, this Fano resonance produced stronger optical absorption and a narrower bandwidth. Angle-resolved spectral analysis is powerful to study the behavior of optical modes[47]. We measured the angle-resolved thermal emission spectra. The results showed that Fano resonance could produce thermal emission with a narrow bandwidth and good directive emission angle. Simultaneously, we established a coupled resonance model to calculate the emission properties of this Fano resonance. The calculated and experimental results agree well.

**Structure Design and Mode Analysis**

The study of the modulated thermal emissions due to the coupling between SLR and MR modes with different parity-symmetries was performed using Al-SiN-Al magnetic MTSs. Firstly, we designed the structure and studied the eigenmodes with different parity-symmetries in the MIM MTS using the commercial finite element method software, COMSOL Multiphysics (COMSOL Colt.). Figure 1(a) provides the geometric and material information, and shows a schematic plot of angle-resolved thermal emission configuration as well, where the polarization is selected along x-direction. In simulation, the width of the grating was fixed at d=3.5 µm, and the thickness of the amorphous SiN and Al gratings were t=0.5 µm and h=0.05 µm, respectively. Based on Kirchhoff's law, the absorptivity spectra were looked at in the simulation. To avoid the resonant absorption of phonons in the SiN (phonons resonant absorption of SiN can be referred to supporting information Fig. S1), a wavelength range of 5–10 µm was chosen. The refractive index of SiN film was obtained by means of infrared spectroscopic ellipsometry measurement (supporting information Fig. S1). And that of Ag and Si were referred to database of software.

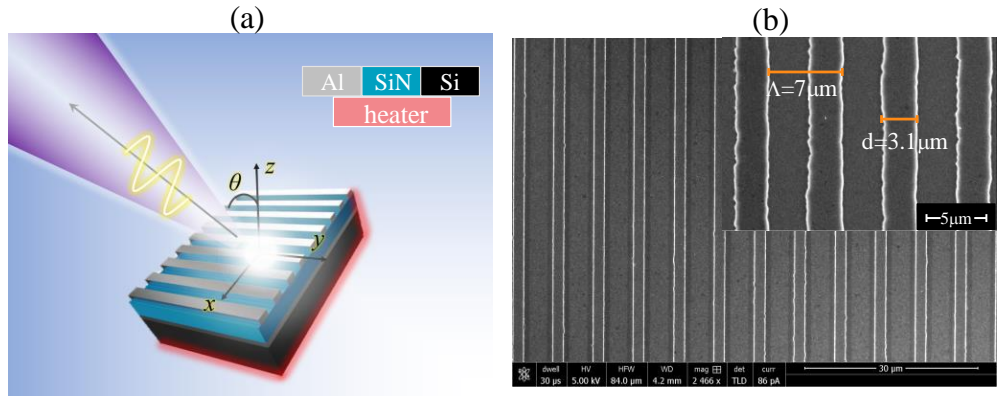

Figure 1. (color online) (a) sketch plot of sandwiched structure, and (b) SEM top view of sample with period Λ=7 µm, grating width d=3.1 µm, as marked in the zoom-in inset picture.

Figure 2(a) and (b) shows the calculated absorption spectra of two samples with different periods (a) Λ=8 µm and (b) Λ=7 µm. Transverse magnetic (TM) polarized light impinged on the samples at normal θ=0° (black line) and θ=20° (red line) incident angles. For clarity, the peaks in Fig. 2(a) and (b) were marked. For the sample with Λ=8 µm, at θ=0° (the black line in Fig. 2(a)), there were three prominent peaks. The first and third peaks located at 9.1 and 5.49 µm, respectively, exhibited Lorentz shapes. Furthermore, one sharp and high peak was superimposed on the first peak located at 8.07 µm, which was close to the lattice period, caused by surface lattice resonance (SLR) due to surface waves diffracted by the period. At the incident angle θ= 20°, the absorption spectrum is shown as a red line in Fig. 2(a) which exhibits different behavior. The SLR peak near 8.07 µm red shifted outside the examined wavelength range. The first and third peaks shifted to longer wavelengths slightly. Furthermore, another Lorentz-shaped peak near 7.2 µm was measured, which did not appear for θ=0° (the black line in Fig. 2(a)). This peak is referred to as the second peak. The first, second, and third peaks corresponded to MR modes, which were insensitive to the lattice. Furthermore, the first and third peaks correspond to bright modes, and the second peak must correspond to a dark mode because it only radiated under oblique incidence[15]. The results of the Λ=7 µm sample are given in Fig. 2(b). At θ=20°, shown as a red line, three MR modes were still detected: a first MR peak at 9.1 µm, a second MR peak at 7.2 µm, and a third MR peak at 5.7 µm. An SLR mode was detected at 9.4 µm. However, at a normal incidence angle (Fig. 2(b) black line), a fascinating phenomenon occurred, as one sharp and asymmetric line peak was detected at 7.14 µm. This is different from the results of the Λ=8 µm sample. At the wavelength of this sharp and asymmetric peak, the second dark MR mode and SLR mode of the sample overlapped. Thus, we preliminarily ascribed it to a Fano resonance mode, which arose due to the strong coupling between the second dark mode and SLR mode. However, no obvious coupling effect between the SLR mode and first MR bright mode could be discerned in Fig. 2(a) and (b) when the first MR bright mode and SLR mode overlapped.

Based on the absorption spectra analysis above, different far-field radiation phenomena occur when the SLR mode couples with the first MR bright mode and second MR dark mode. To exploit this physical mechanism, the spatial distributions of the MR and SLR eigenmodes and the coupled Fano resonance mode were simulated using COMSOL Multiphysics. In the simulations, periodic boundary conditions were employed. Fig.2 (c)–(f) shows the spatial distribution of the vector electric field $\vec{E}_y$ of the (c) third MR, (d) second, and (e) first MR modes, corresponding to the peaks of the red line shown in Fig. 2(a) for Λ=8 μm, located near 5.8, 7.2, and 8.987 μm, respectively. The (e) SLR mode corresponds to the peak of the black line near 8.07 μm in Fig. 2(a). For the MR modes shown in Fig. 2(c)–(e), the electric fields were mainly localized in the dielectric layer, and the electric field of the SLR mode shown in Fig. 2(f) was scattered in free space. Furthermore, the MR spatial symmetry properties can be determined from the vector properties of the electric fields $\vec{E}_y$ indicated by the colored arrows in Fig. 2(c)–(e), where the (a) third and (c) first MR modes were antisymmetric, while the (b) second MR mode was symmetric. Moreover, it was inferred that antiparallel electric circumflux and MR modes occurred[15]. One circumflux is shown in Fig. 2(e) and two pair parallel ones are shown in Fig. 2(a). However, in Fig. 2(b), due to the symmetric electric field distribution, the two antiparallel circumfluxes can be discerned from the weak localized electric field, which would lead to a net zero field located in the dielectric[15]. As a result, the field intensity of the second MR mode (Fig. 2(d)) was much weaker than those of third (Fig. 2(c)) and first (Fig. 2(e)). Clearly, the field parity-symmetric properties play an important role in MR radiation states, the first and third MR modes were parity-antisymmetric bright modes, and the second MR was a parity-symmetry dark mode. Correspondingly, the prominent first and third absorption peaks were revealed in the spectra obtained at normal incidence shown as black lines in in Fig. 2(a) and (c). There was no radiation of the second MR mode at 7.2 μm under normal incidence, as shown by the black line in Fig. 2(a). However, when the second MR dark mode spectrally overlapped with the SLR bright mode, Fano resonance occurred, as shown by the sharp and asymmetric second peak of the black line in Fig. 2(b). The corresponding electric field, shown in Fig. 2(g), was strongly parity-symmetric distributed in the dielectric and scattered into free space.

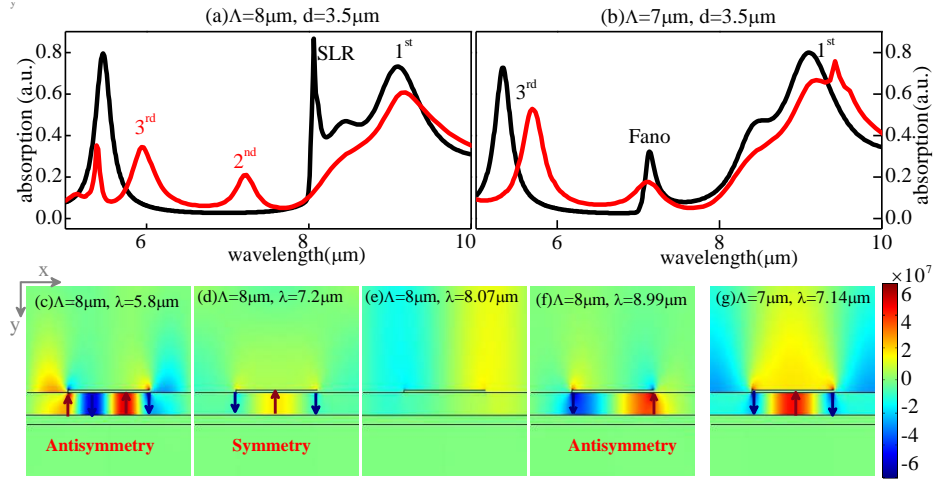

Figure 2. Absorption spectra of samples with period (a) Λ=8 µm and (b) Λ=7 µm under incident angles of θ=0° (the black line) and θ=20° (the red line). Spatial distributions of the vector electric fields $\vec{E}_y$ in the xy-plane of the (c) third, (d) second, (e) and first the MR eigenmodes ( the red line in (a)); the (f) SLR mode (black line in (a)); and (g) the coupled Fano resonance (the black line in (b)). The corresponded wavelengths and lattice of these eigenmodes are marked out. The arrows denote the vector properties of the fields.

As discussed above, the Fano resonance produced a narrow and asymmetric absorption peak. As well known, the spectral behaviors of optical resonance modes are characterized by quality factor (Q-factor). It is defined as Q =$\lambda/(2\Delta\lambda)$[48,49], where λ is the central wavelength of optical resonances, Δλ as the half bandwidth. Q-factor of Fano resonance in Fig. 2(b) (the black line) near λ= 7.14 µm is about 60.4 (detailed calculation method can be referred to Supporting information Fig. S3), which is more than eight-fold of that of magnetic resonance with 7.72 in Fig. 2(b) (the red line) near λ= 7.1 µm. Based on Kirchhoff's law, the thermal emission of the Fano resonance can obviously produce a sharp emission spectrum, which can be used as a narrow band thermal emitter.

**Result and Discussion**

To demonstrate the selective emitter properties achieved by the coupling between SLR and MR with different parity-symmetry, Al/SiN/Al magnetic MTSs were fabricated and studied by fixing the resonant frequency of the SLR with a fixed lattice with Λ=7 μm and altering the resonance frequency of the MRs by changing the grating width d. Firstly, an Al film with a thickness of 0.15 μm was deposited on a cleaned Si wafer by electron beam evaporation. Subsequently, amorphous SiN films (0.5 μm) were synthesized by plasma enhanced chemical evaporation (PECVD) at a temperature of 300℃ on the Al film. Al gratings were fabricated atop the amorphous SiN films using ultra-violet photolithography and subsequent Al film deposition and lift-off. We chose the geometric parameters in Fig. 1(a) as thicknesses of the SiN films t=0.5 μm, an Al grating height h=0.05 μm, and a grating period of Λ=7 μm. The grating width was varied by altering the exposure time in the photolithography process. Finally, grating widths of d=1.45, 1.8, 3.1, and 3.45 μm were obtained. Fig. 1(b) shows the morphology of the sample with d=3.1 μm, characterized by scanning electron microscopy (SEM). The inset in Fig. 1(b) gives out the zoomed in picture, for clarity, both lattice and width of grating are marked out. The angle resolved thermal emissions of all the samples were measured using an FTIR spectrometer with one custom-made rotational sample heater (the detailed setup can be referred to supporting information Fig. S2), where the polarization is along x-direction, and the incident angle θ can be tuned in the range of -90°–90° with a minimum step of 0.01°, as referred in Fig.1(a) the schematic plot. The spectrum measurement range was 5–10 μm.

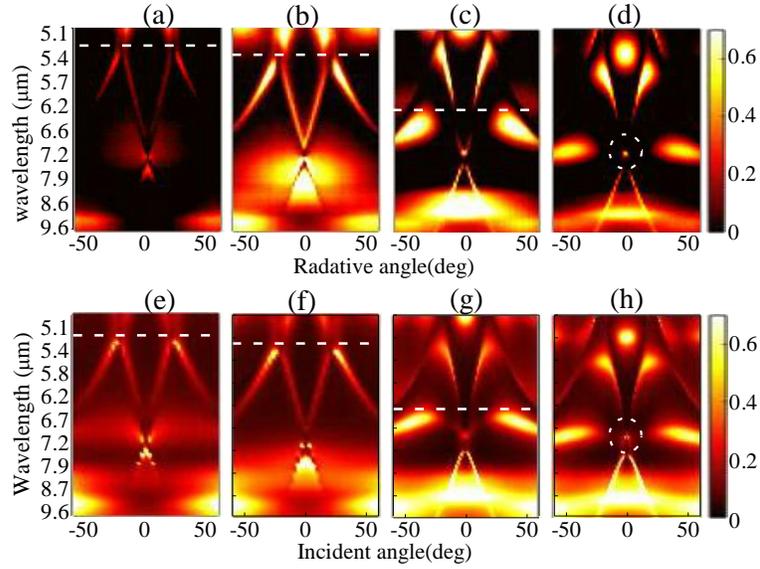

Figure 3. (a)–(d) Measured and (e)–(h) simulated angle-resolved thermal emission spectra of samples with fixed lattice Λ=7 μm and varied grating widths d of (a, e) 1.4, (b, f) 1.8, (c, g) 3.1 and (e, h) 3.45 μm. The second MRs are marked as dashed white line. The spatial isolated spots were highlighted by dashed white circle in (d, h).

The angle resolved thermal emission spectra are shown in Fig. 3 (a)–(d) for the samples with grating widths d=1.4, 1.8, 3.1, and 3.45 μm. In comparison, the simulated absorption spectra of these samples are shown in Fig. 3 (e)–(h).

The angle resolved spectra of the measured thermal emissions exhibited similar properties to those of the simulated spectra. As can be seen from Fig. 3(a)–(c) and (e)–(g), at a normal angle, the bands of the SLR bright mode were folded into one crossing cone near wavelength 7.18 μm. The first and second MR modes did not change with angle but changed with increase grating width d. Clearly, at a normal angle, the angle resolved spectra exhibited strong radiation for the first MR parity-antisymmetric mode, as could be seen in Fig. 3 the broad and bright zone. When d=1.45 µm, the first MR and SLR modes spectrally overlapped near λ= 7.2 µm (a, e). But the first MR parity-antisymmetric bright mode could still be discerned, and no obvious changes occurred. As for the sample with a grating width of d=1.8 µm, the first MR parity-antisymmetric bright mode moved below the cone near λ= 7.9 µm as shown in Fig. 3(b) and (f). With further increases in the grating width, as shown in Fig. 3(c, g) and 3(d, h), the frequency location of first MR parity-antisymmetric bright mode was far from that of SLR. There's no coupling of the first MR and SLR parity-antisymmetric bright mode for all samples. However, for the second MR parity-symmetric mode, weak radiation was exhibited at normal incidence, as marked out by the dashed white line. With width of grating increasing, the frequency location of second MR parity-symmetric dark mode (dashed white line) gradually approached that of the SLR folding cone. At the grating width d=3.45 µm, the wavelength of second MR parity-symmetric dark mode spectrally overlapped with that of the SLR folding cone, and one isolated bright thermal emission spot can be found at the normal incident angle near λ=7.1 µm, as highlighted by the dashed white circle in Fig. 3(d) and (h). According to the above discussion, this isolated bright thermal emission light spot was induced by Fano resonance due to the strong coupling between the second MR parity-symmetric dark mode and the SLR mode.

**Numerical Analysis**

To investigate the coupling physical mechanism of the MR with different parity-symmetric properties with SLR, parity-based coupled mode theory is developed, and the above experimental results can be well explained. When harmonic light illuminates a dipole $\mu$, the oscillation equation is given as follows:

$$\ddot{\vec{\mu}} + \gamma\dot{\vec{\mu}} + \omega_0^2\vec{\mu} = gAe^{i\omega t}, \tag{1}$$

where $\omega_0$ is the eigenfrequency of the resonator, g is the coupling factor of the resonator to the external field, which could be determined by the intrinsic properties of the resonator, such as the material, shape factor, and more importantly, the parity. Here, we take g as the mode parity factor, and, in this one-dimensional periodic system, it is defined as

$$g = p(x) - p(-x),  \quad (2)$$

Where $p(x)$ is the field spatial distribution function in the x-direction. For the parity-antisymmetric bright mode, g, and for the parity-symmetric dark mode, g=0. We can mimic the coupling efficiency between the optical modes and light through g. In a system with two optical modes, the state evolution can be written as follows:[50]

$$\ddot{\vec{\mu}}_1 + \gamma_1 \dot{\vec{\mu}}_1 + (\omega_0 + \delta)^2 \vec{\mu}_1 + \kappa \vec{\mu}_2 = g_1 A e^{i\omega t}$$
$$\ddot{\vec{\mu}}_2 + \gamma_2 \dot{\vec{\mu}}_2 + \omega_0^2 \vec{\mu}_2 - \kappa \vec{\mu}_1 = g_2 A e^{i\omega t} \quad (3)$$

Here, $\mu_1, \mu_2$ are the charge displacements corresponding to different optical modes, $g_1, g_2$ are the parity factors that indicate the coupling between the optical modes and incident field, $\kappa$ is the coupling factor between the two optical modes,[40,50] $\delta$ is the frequency delay between eigenfrequencies of the two optical modes, and $\gamma_1, \gamma_2$ are the dissipation losses of the two optical modes. With the normalized $A=1$, the amplitude of the charge displacement can be described as follows:

$$\begin{pmatrix} \tilde{\mu}_1 \\ \tilde{\mu}_2 \end{pmatrix} = - \begin{pmatrix} \omega_0 - i\gamma_1 & \kappa \\ \kappa & \delta + \omega_0 - i\gamma_2 \end{pmatrix}^{-1} \begin{pmatrix} g_1 \\ g_2 \end{pmatrix} \tilde{E}_0. \quad (4)$$

By solving Eq. (4), we obtained the amplitude of the charge displacement of the two optical modes as follows:

$$\tilde{\mu}_1 = \frac{-\kappa g_2 + g_1(\delta + \omega_0 - i\gamma_2)}{\kappa^2 - (\delta + \omega_0 - i\gamma_2)(\omega_0 - i\gamma_1)} \quad (5)$$

$$\tilde{\mu}_2 = \frac{g_2(\omega_0 - i\gamma_1) - \kappa g_1}{\kappa^2 - (\delta + \omega_0 - i\gamma_2)(\omega_0 - i\gamma_1)} \qquad (6)$$

Finally, the energy dissipation in the system can be calculated as[50]

$$P = \frac{\omega^2}{2}\left(\gamma_1|\tilde{\mu}_1(\omega)|^2 + \gamma_2|\tilde{\mu}_2(\omega)|^2\right) \qquad (7)$$

We numerically calculated the absorption spectra using Eqs. (5)–(7). First, the parameters $\gamma_1, \gamma_2$ and $\omega_0$ of the SLR and MR modes were obtained from Lorentz line shape fitting. For the mode parity parameters, $g_1, g_2$, the symmetric properties of the electric field $\vec{E}_y$ distribution in Fig. 2 indicated that, for the parity-antisymmetric bright mode of the (a) first and (c) third MR modes and the (d) SLR mode, $p(-x) = -p(x)$, whereas for the parity-symmetry dark mode of the second MR mode, $p(-x) = p(x)$. Thus,

$$g_1 = p(x) - p(-x) = 2p(x)$$

$$g_2 = p(x) - p(-x) = 0$$

To balance the energy, we set the value of g in the range $[0, 0.5]$.

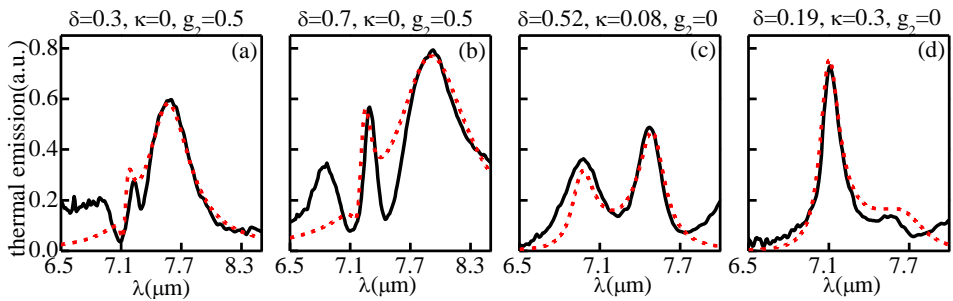

Figure 4. Numerical calculated absorptivity (dashed red line) and measured thermal emission (black line) spectra versus wavelength of samples with $d=$ (a) 1.45, (b) 1.8, (c) 3.1, and (d) 3.45μm.

The calculated absorption (dashed red line) spectra and measured thermal emission (black line) spectra versus wavelength are shown in Fig. 4. The calculated absorptivity using parity-based coupled mode theory agreed well with the measured thermal emission. At the location near $\lambda \approx \Lambda \approx 7$ µm, due to the different interactions between the MRs with different parities and the SLR, the spectra provide new information. For the samples with (a) $d$=1.45 µm and (b) $d=1.8$ µm, the parity factor of the SLR and the first MR modes were $g_1$=0.5 and $g_2$=0.5, and the spectra in Fig. 4(a) and (b) exhibited similar line shapes, with one narrow peak (SLR) superimposed on the broad first MR peak. Even for the sample with $d$=1.45 µm in Fig. 4(a), where the first MR parity-antisymmetric bright mode exhibited significant frequency overlapping with the SLR mode, no coupling occurred with $\kappa = 0$. As the grating width increased, the first MR moved away from the SLR, the second MR parity-symmetry dark mode approached the SLR, and the parity factors were $g_2$=0 and $g_1$=0.5 for the second MR and SLR modes, respectively. For the sample with $d=3.1$ µm shown in Fig. 4(c), the spectrum exhibited a different behaviour than that in Fig. 4(a) and (b). Two similar peaks at λ=6.95 and 7.47µm appeared in the spectral line. Clearly, with $g_1$=0.5 and $g_2$=0, the coupling properties were different from those of the first MR and SLR modes, although the frequency delay was larger as $\delta$=0.52 µm, and the coupling was still weak with $\kappa = 0.08$, yielding new optical modes. Further increasing the grating width to $d$=3.45 µm, an intriguing property is highlighted in Fig. 4(d): one narrow, sharp, and asymmetric line-shaped peak located at λ=7.115 µm appeared, which also corresponded to the Fano resonance line shape due to the strong coupling of the second MR antisymmetric dark mode and SLR mode with $\kappa = 0.3$. Comparing thermal emission spectrum of coupled one in figure 4(d, the black line) and figure 4(a-c, black lines), there's no other resonance near the thermal emission peak near λ=7.115 µm in figure 4(d).

**Conclusion**

Meta-surfaces of nano-sandwich optical absorbers Al/SiN/Al were used to control the thermal emission in the infrared range. Analysis of the MR modes and their coupling effects with the SLR mode was performed using COMSOL Multiphysics. No coupling occurred between the first MR parity-anti-symmetric bright mode and the SLR mode, while Fano resonance was induced by the coupling between the second parity-symmetric dark mode and the SLR mode. A series of samples with varied grating widths were fabricated, and their thermal emissions were experimentally investigated through the angle-resolved thermal emission spectra of the samples. For the sample with $\Lambda=7$ μm and $d=3.45$ μm, one isolated point near the normal incidence angle was observed in the angle-resolved thermal emission spectrum. Numerical analysis based on coupled mode theory revealed that this intriguing phenomenon was caused by Fano resonance due to the strong coupling between the second MR parity-symmetric dark mode and the SLR mode when their wavelengths overlap. As denoted in Fig. 1(a), this coupling only occurs in the xz-plane. In yz-plane, the thermal emission of SLR has different behaviour due to the anisotropic properties of grating (supplementary information in Fig. s5), the coupling with MR would not happen.

**Supplementary information**

We presented the optical properties of SiN film used in simulation, which comes from the infrared ellipsometry measurement. The schematic plot of thermal emission measurement used in this manuscript are given out in the supplementary information. In addition, the Q factor of Fano resonance and purely magnetic resonance mode are calculated, and high Q factor of Fano resonance are demonstrated. At last, the anisotropic properties of this parity symmetric coupling are illustrated, where the coupling of SLR and parity-symmetric MR occurs in the xz-plane.

**Author information**

**Corresponding Authors**

*E-mail: liuhui@nju.edu.cn

*E-mail: xzhangqf@qfnu.edu.cn

**Notes**

The authors declare no competing financial interest

**Funding Sources**

The National Natural Science Foundation of China (Nos. 11321063, 61425018, and 11374151) and the National Key Projects supported this work for Basic Research in China (Nos. 2012CB933501 and 2012CB921500).

# Supporting information

# Controlling Thermal Emission by Parity-symmetric Fano Resonance of Optical Absorbers in Metasurfaces

*Xia. Zhang[1,2]\*, Zhenguo. Zhang[1], Qiang. Wang[1], Shining. Zhu[1] and Hui. Liu[1]\**

[1]National laboratory of solid state microstructures & school of physics, Collaborative Innovation Center of Advanced Microstructures, Nanjing University, Nanjing 210093, China

[2]School of Physics and Physical Engineering, Shandong Provincial Key Laboratory of Laser Polarization and Information Technology, Qufu Normal University, Qufu 273165, China

1. **Optical properties of SiN used in simution**

In simulation, the optical properties parameters of SiN film on metal substrate play important role. We fabricated SiN film with thickness of 500 nm on Al film (150 nm), and measured by the infrared ellipsometry at angle θ= 65 deg, and the optical property parameters of SiN were fitted using Lorentz model. Fig. S1 shows optical properties used in simulation. Clearly, there's one broad absorption peak ranged from near λ= 10 to 15 μm comes from the phonon of SiN. To avoid the absorption effect of phonons in the SiN, a wavelength range of 5–10 μm was chosen in the main text.

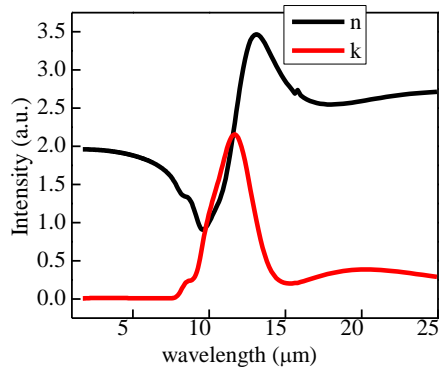

Figure S1. (color online). Refractive index (black curve) and absorption coefficient (red curve) of SiN film on Al fabricated by PECVD.

## 2．angle-resolved thermal emission measurement setup

The angle resolved thermal emission spectra shown in the main text in Fig. 3 (a)–(d) were measured on homemade setup as Fig. S2. The sample was attached on an electrical heater joined onto rotatable stage, we can change the emission angle through rotating the sample. The emitted signal was spatial selected by one slit to low the stray light, and was polarized with one polarizer plate. And then, it was collected by the detector inside the FTIR spectrometer, as can be seen in Fig. 2S. The collected data was referenced to that of blackbody.

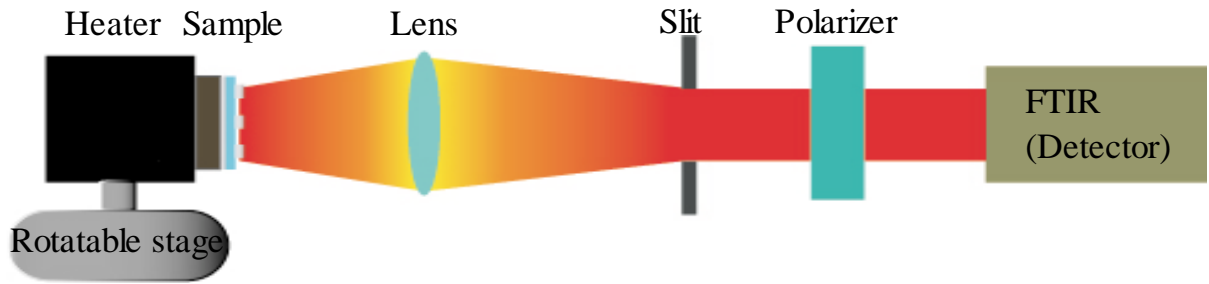

Fig. S2 Schematic plot of thermal emission spectra measurement setup.

## 3. Q-factor analysis

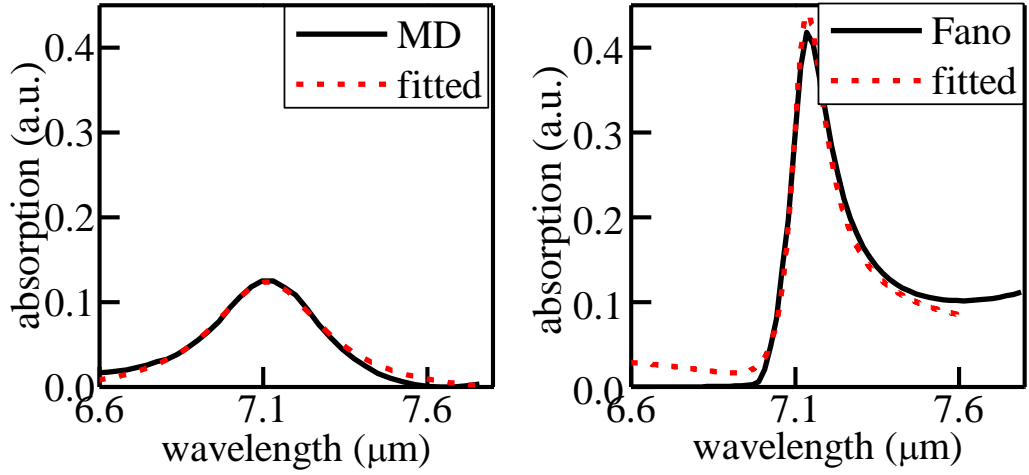

Fig. S3 Q-factor fitting results (a) magnetic resonance, (b) Fano resonance.

The spectral behaviors of optical resonance modes are characterized by quality factor (Q-factor). It is defined as

$$Q = \lambda_0/(2\Delta\lambda) \quad (1)$$

where $\lambda$ is the central wavelength of optical resonances, $\Delta\lambda$ as the half bandwidth. To obtain these parameters, the Lorentz and Fano fitting were performed for simulated magnetic resonance and Fano resonance in Fig. 2 in the main text, respectively. Fig. S3 gave out the fitting results, where the black lines were the simulation results and the red dashed lines were fitting results.

The Lorentz fitting for simulated magnetic resonance was performed based on formula as

$$P_{MR} = a + A \times \frac{w}{4 \times (\omega - \omega_0)^2 + w^2} \quad (2)$$

Where $\omega_0$ was the central frequency, $w$ was bandwidth, $a$ and $A$ were constant, $\omega$ ranged in [6.6, 7.6] in the unit of microns. After iteration fitting, we obtained the results as denoted in Fig. S3(a), where $\omega_0 = 7.11$ and $w = 0.46$, both in the unit of microns, were corresponded to $\lambda_0$ and $\Delta\lambda$ in formula (1). The Q-factor was about 7.72.

The Fano fitting for simulated absorption peak 2 in Fig. 2(the black line) was performed based on formula as

$$P_{Fano} = a + jb + A \times \frac{w}{(\omega - \omega_0 + iw)^2} \quad (3)$$

Besides parameters $\omega_0$, $w$, and $A$, here, the offset parameters were complex as $(a + jb)$. The fitting result was given in Fig. S3(b). It agrees the simulated one well, the parameters were $\omega_0 = 7.13$ and $w = 0.059$, and the Q-factor amounted to 60.4, which is about eight-fold that of magnetic resonance.

**4. Comparison of thermal emission in xz-plane and yz-plane.**

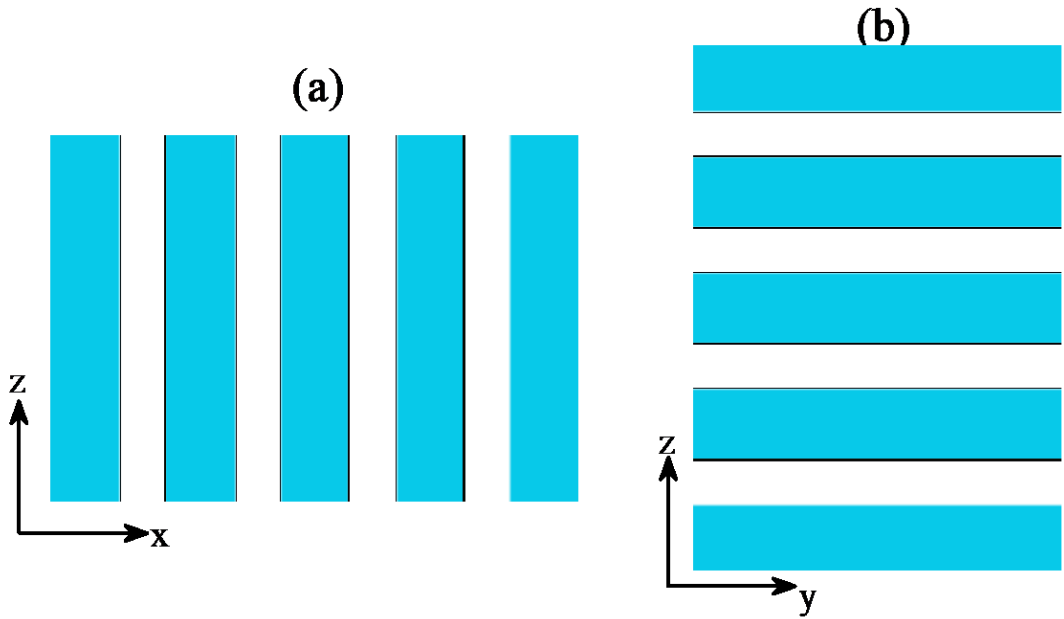

Fig. S4 the schematic plot of angle resolved thermal emission measurement (a) in xz-plane and (b) in yz-plane.

Fig. s4 present the schematic plot of angle resolved thermal emission measurement both in xz-plane (a) and in yz-plane. The measurement was performed in Al/SiO$_2$/Al grating system, where the period $\Lambda = 6$ μm, Al film with a thickness of 0.15 μm was deposited on a cleaned Si

wafer by electron beam evaporation, amorphous $SiO_2$ films (0.5 µm) were synthesized by plasma enhanced chemical evaporation (PECVD) at a temperature of 300℃ on the Al film, the thickness of Al gratings h=0.05 µm were fabricated atop the amorphous SiN films using ultra-violet photolithography and subsequent Al film deposition and lift-off. Angle resolved thermal emission of sample with grating width h=4.1 µm is given out in Fig. s5 both in xz-plane (left) and yz-plane (right). To discern the SLR and MR mode, the thermal emission in the xz-plane are also marked out. Here, the 4th MR (dashed red line) is dark mode, which can't be discerned in the thermal emission in yz-plane (right) even in large incident angle. However, the 3rd MR (dashed yellow line) bitght can be clearly discerned both in xz-plane and yz-plane, it varies slightly with the incident angle. The thermal emissions of SLR in xz-plane and yz-plane (dashed dark blue line) demonstrate completely different behavior. It deviates with increasing the incident angle in yz-plane due to the projection effect.

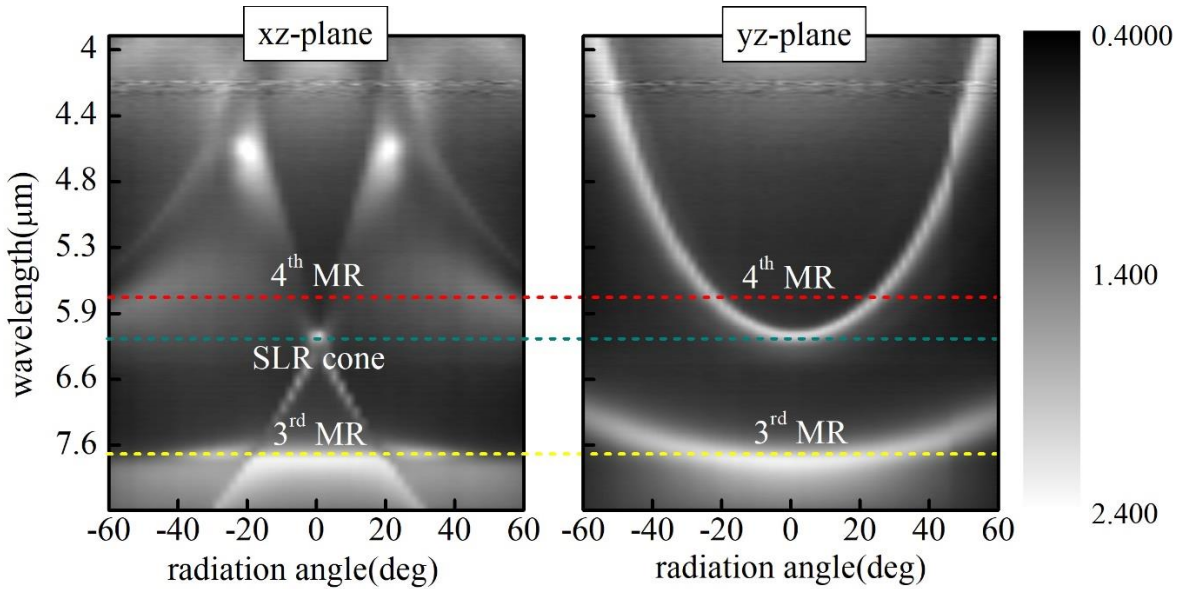

Fig. s5 the angle-resolved thermal emission spectra in the xz-plane and yz-plane of samples Al/$SiO_2$/Al grating with fixed lattice Λ=6 µm, d=4.1 µm.